\documentclass[a4paper]{jpconf}
\usepackage{graphicx}
\usepackage{hyperref}
\usepackage[section]{placeins}
\begin{document} 

\title{Investigating a New Approach to Quasinormal Modes: Physics-Informed Neural Networks}
\author{A M Ncube, G E Harmsen and A S Cornell}
\address{Department of Physics, University of Johannesburg, PO Box 524, Auckland Park 2006, South Africa.}
\ead{ncubeanele4@gmail.com, gerhard.harmsen5@gmail.com, alanc76@gmail.com}

\begin{abstract}
Physics-informed neural networks (PINNs) hold the potential for supplementing the existing set of techniques for solving differential equations that emerge in the study of black hole quasinormal modes. The present research investigated them by studying black hole perturbation equations with known analytical solutions and thus could be framed as inverse problems in PINNs. Our main goal was to test the accuracy of PINNs in computing unknown quasinormal frequencies within the differential equations.  The black hole perturbation scenarios that we considered included near extremal Schwarzschild-de Sitter and Reissner-Nordstr\"{o}m-de Sitter black holes, and a toy problem resembling them. For these cases, it was shown that PINNs could compute the QNFs with up to 4 digit decimal accuracy for the lowest multipole number, $l$, and lowest mode number, $n$.
\end{abstract}

\section{Introduction}

A black hole responds to test field perturbations by producing quasinormal modes (QNMs) that dissipate over time as the black hole returns to its initial state of equilibrium \cite{DeyChakrabarti2019}. Similar to the well-known phenomenon of standing waves, these QNMs oscillate with a set of discrete resonant frequencies, known as quasinormal frequencies (QNFs), which are represented by complex-valued scalars. The real part is the physical oscillation frequency of the QNM and the imaginary part is proportional to its damping rate. Since these frequencies depend only on the physical parameters of their source, they are ideal for probing the characteristics of black holes \cite{KonoplyaZhidenko2011, IyerWill1987}.

Mathematically, black hole QNMs are solutions to second-order differential equations describing the space-time evolution of perturbing fields in the vicinity of a black hole. Given that many of these equations are analytically intractable, several approximation techniques have arisen to solve them over the past decades. These methods have been outlined concisely in recent review articles such as Ref.~\cite{KonoplyaZhidenko2011} and include a modified form of the asymptotic iteration method \cite{Choetal2012} initiated by one of the authors of this proceeding.

This proceeding, however, is focused on investigating physics-informed neural networks (PINNs) to determine whether they could compute black hole QNMs. PINNs were recently introduced in the literature and are derived from traditional artificial neural networks. These machine learning algorithms are known for their capabilities as universal function approximators \cite{Raissietal2019}. We have considered them in our study of black hole perturbation equations because of their capacity to solve differential equations. The key to their operation is the inclusion of physical constraints in the loss function, which is used to steer them towards accurate solutions. Note that PINNs have been shown to be more effective at solving inverse problems than forward-modelling problems where they are currently surpassed by existing numerical mesh-based techniques \cite{Karniadarkisetal2021}. Inverse problems are differential equations with unknown parameters that can be determined from observational data of the solution. On the other hand, forward problems are well-posed differential equations that require only the boundary and/or initial conditions of a given physics problem for the PINNs to solve them.

\section{Perturbations of spherically symmetric black holes} 

To arrive at the differential equations of black hole perturbations, we begin by considering the space-times of spherically symmetric black holes given generally as \cite{CardosoLemos2003}:
  
\begin{equation}
ds^2 = -f(r) dt^2 + f(r)^{-1} dr^2 + r^2(d\theta^2 + \mathrm{sin}^2\theta d\phi^2).
\end{equation}

Here $r$ is the radial distance from the centre of the black hole and $f(r)$ is a metric function which, for Schwarzschild-de Sitter (SdS) and Reissner-Nordstr\"{o}m-de Sitter (RNdS) black holes is \cite{CardosoLemos2003, Molina2003}:

\begin{equation}
f(r) = 1 - \frac{2M}{r} + \frac{Q^2}{r^2} - \frac{\Lambda r^3}{3},
\end{equation}

where $M$ and $Q$ are the mass and electric charge of the black hole in geometrical units, and $\Lambda$ is the cosmological constant. The SdS black hole has no electric charge; thus, the term $Q$ in the metric function vanishes. For de-Sitter black holes, the cosmological constant is positive as these space-times have positive curvature as $r \rightarrow +\infty$. 

Building on these black hole space-times, the perturbation equations for given perturbing fields are derived from the equations of motion of these fields. For example, the perturbation equation for an electromagnetic test field in the vicinity of a black hole is obtained from the source-free Gauss-Amp\`{e}re law of Maxwell's equations \cite{dInverno1992, Carroll1997}:

\begin{equation}
F^{\mu\nu}_{\ \ ;\nu} = \frac{1}{\sqrt{|g|}} \partial_{\nu}\left(\sqrt{|g|} F^{\mu\nu} \right) = 0.
\end{equation} 

Here $F^{\mu\nu}$ is the electromagnetic field tensor and $|g|$ is the determinant of the metric tensor of the curved space-time around the black hole. When we consider metric tensors of spherically symmetric black holes, the equations of motion for perturbing fields simplify to “Schr\"{o}dinger-like” black hole perturbation equations given generally as \cite{CardosoLemos2003}:

\begin{equation}
\frac{d^2 \psi(r)}{d x^2} +\ (\omega^2 - V(r))\psi(r) = 0,
\end{equation}

where $x$ is the “tortoise” coordinate that is related to $r$ by the relation: $dr/dx = f(r)$ \cite{KonoplyaZhidenko2011}. The black hole effective potential, $V(r)$, for massless scalar perturbations of SdS and RNdS black holes is given as \cite{Choetal2012, Molina2003}:

\begin{equation}
V(r) = f\left[ \frac{l(l + 1)}{r^2} + \left(\frac{2M}{r^3} - \frac{2Q^2}{r^4}- \frac{2\Lambda}{3} \right)\right],
\end{equation}

where $l$ is the multipole number and $Q = 0$ for a SdS black hole. When black hole QNMs have negative temporal dependence they satisfy the boundary conditions given by $\psi(x) \sim \mathrm{exp}(\pm i\omega x)$ as $x \rightarrow \pm \infty$. Physically, this implies that the QNMs can only radiate inward at the horizon and outward at spatial infinity \cite{Leaver1985, Stein2019} at a time $t > 0$.

Due to the nature of the effective potential $V(r)$, equation (4) is generally difficult to solve analytically except for some special cases such as those considered here for simplicity. These space-times are the SdS and RNdS black holes in the near extremal limit where the cosmological horizons of the space-times are very close (in the $r$ coordinate) to their event horizons.

Ref.~\cite{Molina2003} demonstrated that for any de-Sitter black hole space-time in the near extremal limit, the metric function $f(r)$ can be written as:

\begin{equation}
f(r(x)) =  \frac{(r_2 - r_1)\kappa_1}{2 \mathrm{cosh}^2{\kappa_1 x}} + \mathcal{O}(\delta^3),  
\end{equation}

where $\delta = (r_2 - r_1)/r_1$, $\kappa_1$ is the surface gravity at the horizon, $r_1$ and $r_{2}$ are two consecutive positive roots of $f(r)$,  and $x$ is the tortoise coordinate whose domain lies within ($r_1$, $r_2$). For both the SdS and RNdS, $r_1$ and $r_2$ are the event and cosmological horizons, respectively, with $r_2 > r_1$. 
Consequently, in the near extremal limit the two black holes share the same mathematical expression for the metric function, which in turn results in a single expression for the effective potential valid for both SdS and RNdS black holes \cite{CardosoLemos2003, Molina2003}:

\begin{equation}
V(x) = \frac{V_0}{\cosh^2(\kappa_{_b}x)}.
\end{equation}

 Here $V_0 = \kappa_b^2 l(l+1)$ for massless scalar perturbations and $\kappa_b$ is the surface gravity associated with the black hole horizon. With the effective potential in this form, the black hole perturbation equations yield analytic solutions.

In the case of massless scalar perturbations of SdS and RNdS black holes, the exact analytic expressions for the QNMs (denoted by $\psi(x)$) and QNFs (denoted by $\omega$) are, respectively \cite{CardosoLemos2003, Molina2003, FerrariMashhoon1984}:

\begin{equation}
\psi(x) =[\xi(\xi -1)]^{i\omega/2\kappa_{_b}} \cdot  {}_2F_1\left(1 + \beta + i\frac{\omega}{\kappa_{_b}}, - \beta + i\frac{\omega}{\kappa_{_b}}; 1 + i\frac{\omega}{\kappa_{_b}}; \xi\right),
\end{equation}

\begin{equation}
\frac{\omega}{\kappa_{b}} = \sqrt{\left(l(l + 1) - \frac{1}{4}\right)} - i\left(n + \frac{1}{2}\right),\quad n = 0, 1, 2, ... 
\end{equation}

where $\xi^{-1} = 1 + \exp{(-2\kappa_{_b}x)}$ and $\beta = -1/2 + (1/4 - V_0/\kappa_{_b}^2)^{1/2}$. 

With these exact solutions, a dataset was generated to provide “observational biases” \cite{Karniadarkisetal2021} for training the PINNs to solve the perturbation equations given as inverse problems.

Another problem similar to the perturbations of SdS and RNdS black holes, which was discussed in Ref.~\cite{Choetal2012}, involves a simpler form of the inverted P\"{o}schl-Teller potential given as:

\begin{equation}
V(x) = \frac{1}{2\mathrm{cosh}^2(x)}.
\end{equation}

The aforementioned boundary conditions of black hole QNMs are also satisfied by the QNMs of this problem. In this case, the exact form of the QNMs, obtained in Ref.~\cite{ChoHo2007} using quasi-exactly solvable theory, are:

\begin{equation}
\psi(x) = (\mathrm{cosh}(x))^{(i + 1)/2}\chi_n(\mathrm{sinh}(x)), 
\end{equation}
\begin{equation}
\omega_n = \pm \frac{1}{2} - i(n + \frac{1}{2}),
\end{equation}

where $\chi_n$ is a polynomial of degree $n$ in $\mathrm{sinh}(x)$. This problem and the previously mentioned one pertaining to near extremal SdS and RNdS black holes were used to test the accuracy of PINNs in computing QNFs.

\section{Computing QNMs with PINNs} 

The PINN algorithm, and the steps taken to implement it on black hole perturbation equations, have two components \cite{Karniadarkisetal2021}:

\begin{enumerate}
    \item A physics-uninformed artificial neural network of a particular architecture, such as a feed-forward neural network (FNN). It represents the neural network solution to the differential equation (figure 1 (left)).
   \item A \textit{loss-function} that measures the deviation of the neural network solution from the physical constraints of the problem (figure 1 (right)). During the training phase, these constraints force the PINN solution to satisfy the boundary and initial conditions of the problem in addition to a labelled training dataset. At the same time, the neural network is forced to minimise a residual associated with the differential equation.
\end{enumerate}

\begin{figure}[ht!]
\begin{center}
\includegraphics[width=0.9\linewidth]{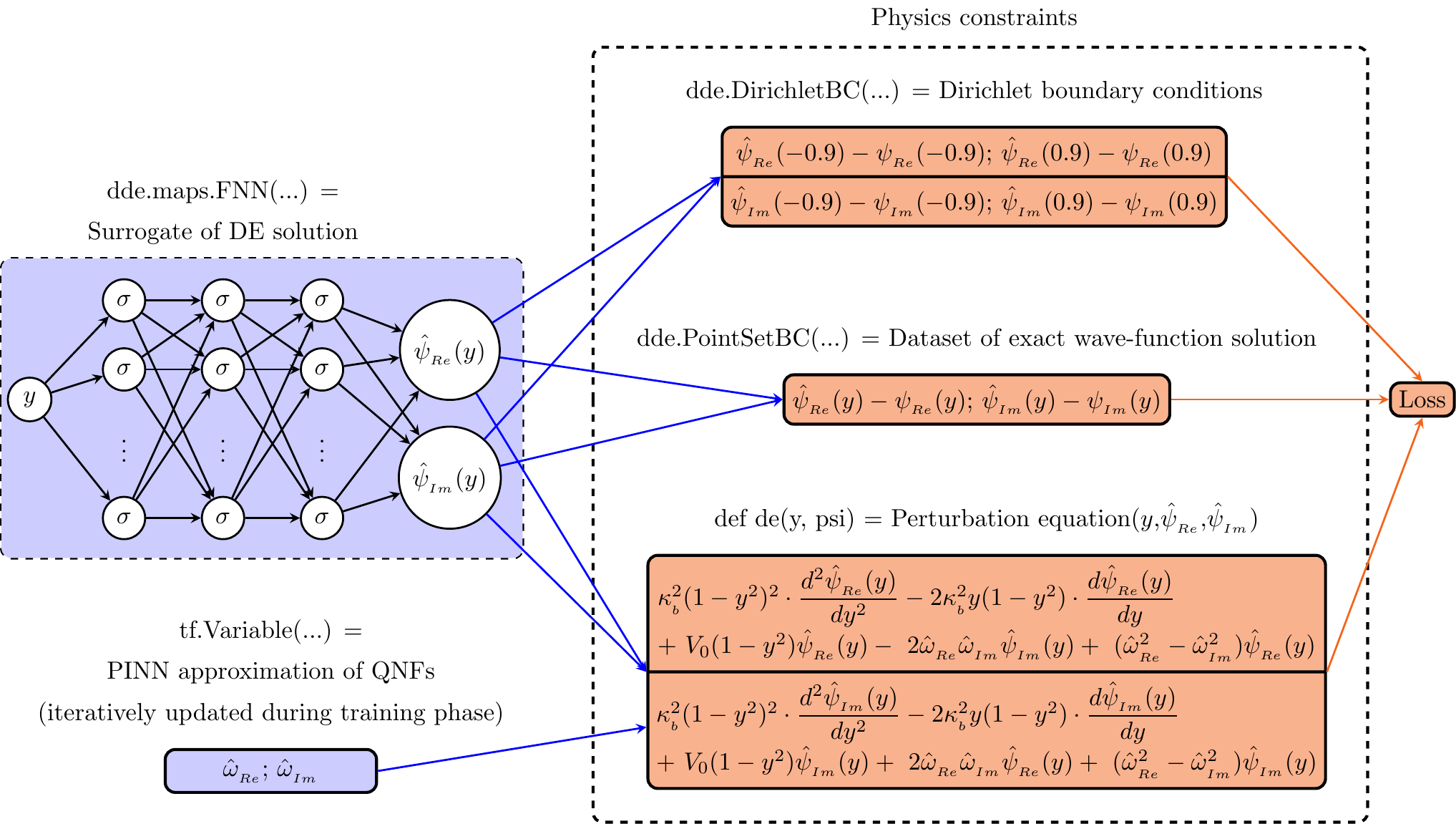}
\small\caption{\  A schematic of the PINN set-up that was built to solve the perturbation equation with the effective potential given by an inverted P\"{o}schl-Teller potential.}    
\end{center}
\end{figure}

The algorithm for building PINN models is outlined in figure 2. Initially, the physics equations governing the problem are all defined and included as arguments for \texttt{deepxde.data.PDE(...)}. The neural network part is set up by defining the specific neural network architecture, the number of hidden layers, nodes per layer and the non-linear activation function used in the nodes. The PINN model is the sum of the neural network (denoted by \texttt{net} in figure 2) and the physical constraints (denoted by \texttt{data} in figure 2) set as arguments for \texttt{deepxde.Model}. The parameters for training, such as the specific choice for the optimiser, are set up using the function \texttt{Model.Compile(...)}. Thereafter, the PINN model can be run for a specified number of training epochs using the command \texttt{Model.Train(...)}.

\begin{figure}[ht!]
\begin{center}
\includegraphics[width=0.7\linewidth]{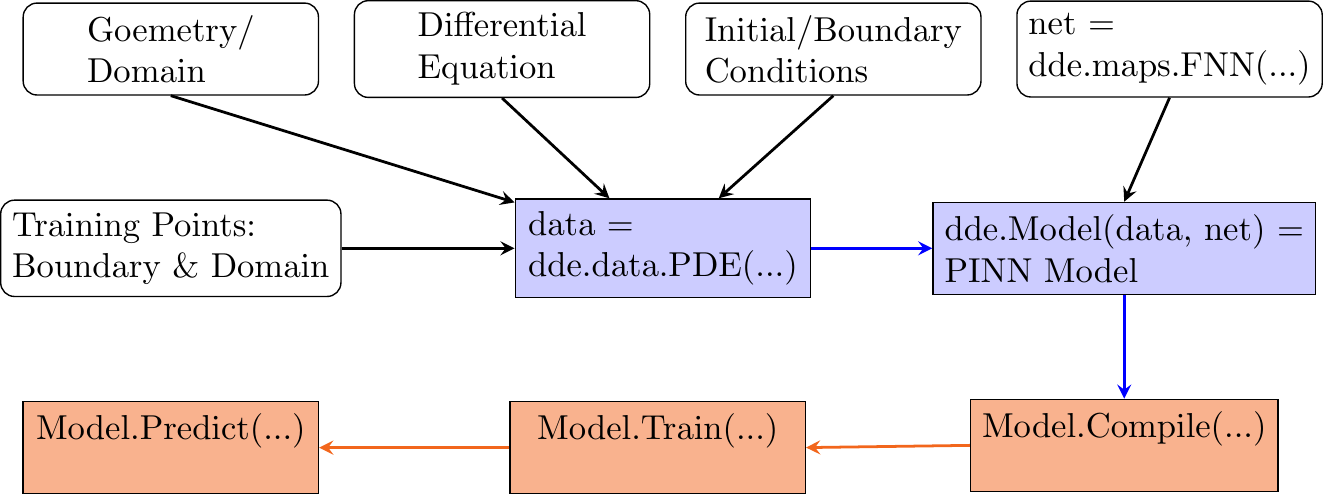}
\caption{\ The PINN algorithm in DeepXDE \cite{Luetal2019}.}
\end{center}
\end{figure}

The same steps were followed in customising PINNs to solve our inverse problem involving massless scalar perturbations of SdS and RNdS black holes with an inverted P\"{o}schl-Teller potential. Since the domain of the problem in the original form is infinite, a new co-ordinate, $y=\mathrm{tanh}(x)$, was used to give a finite domain (-1,1) which is easier to implement in code. Consequently, the QNM boundary conditions become $\psi(y) = \left(1 \mp y\right)^{-i\omega/2}$, as $y \rightarrow \pm 1$. Thus, in terms of  $y$, the perturbation equations for the near extremal SdS and RNdS black holes are:

\begin{equation}
 \kappa_{_b}^2 (1 - y^2)^2 \cdot \frac{d^2\psi(y)}{dy^2} - 2 \kappa_{_b}^2 y (1 - y^2) \cdot \frac{d\psi(y)}{dy} + \left[ \omega^2 - V_0 (1 - y^2)\right]\psi(y) = 0.
\end{equation}

This equation takes the same form in the case of the problem with an effective potential given by equation (10), except that $\kappa_{_b} = 1$ and $V_0 = 1/2$. For easier implementation within the DeepXDE package the equations were split into real and imaginary parts. Figure 1 illustrates the PINN that was used to solve the perturbation equations. 

The training hyperparameters used in all our computations were the following: FNN with 3 hidden layers, 20 nodes per layer; \textit{tanh} as a non-linear activation function; Adam optimiser, which is a standard optimisation method in neural networks \cite{KingmaBa2015}; 20 000 training epochs; training data consisting of 100 domain points, and a dataset with 50 uniformly distributed true values of the QNMs ($\psi(y)$) in the “spatial” domain [-0.9,0.9]. 
 
\section{Results}

The results for the first inverse problem with the SdS and RNdS black holes, and the second inverse problem with an inverted P\"{o}schl-Teller potential given by equation (10), and are listed in tables 1 and 2, respectively. To measure the accuracy of the PINN approximations, we used the L2 relative as a metric \cite{Luetal2019}. For the SdS and RNdS cases, we obtained at most 4 digit decimal accuracy with errors increasing with higher values of $l$. As seen in table 2, the accuracy of the PINN model for the other problem decreased with higher values of $n$.

\begin{table}[ht!]
\caption{\label{Table 1} PINN approximations of the QNFs (in geometrical units) for massless scalar perturbations of SdS and RNdS black holes.}
\begin{center}
\begin{tabular}{cccccccccccr}
\br
  & && \multicolumn{2}{c}{PINN approximation}  && \multicolumn{2}{c}{Exact \cite{CardosoLemos2003, Molina2003}} && $L_2$ \\ 
$n$ & $l$ &&   $\omega_{_{Re}}/\kappa_{_b}$   &   $\omega_{_{Im}}/\kappa_{_b}$  &&   $\omega_{_{Re}}/\kappa_{_b}$    &  $\omega_{_{Im}}/\kappa_{_b}$ &&  relative error \\
\mr
0 & 1 && 1.32287 & -0.49997 && 1.32288 & -0.50000 && 0.00214\%\\ 
 & 2 && 2.39790 & -0.50002 && 2.39792 & -0.50000 && 0.00106\%\\ 
 & 3 && 3.42761 & -0.50032 && 3.42783 & -0.50000 && 0.01105\%\\ 
 & 4 && 4.44096 & -4.98165 && 4.44410 & -0.50000 && 0.08128\%\\
 & 5 && 5.43513 & -0.50716 && 5.45436 & -0.50000 && 0.37463\%\\ 
\br
\end{tabular}
\end{center}
\end{table}

\begin{table}[ht!]
\caption{\label{Table 2} PINN approximations of the QNFs (in geometrical units) for the “Schr\"{o}dinger-like” equation with the simple effective potential given by equation (10).}
\begin{center}
\begin{tabular}{cccccccccccr}
\br
 && \multicolumn{2}{c}{PINN approximation}  && \multicolumn{2}{c}{ Exact \cite{Choetal2012, ChoHo2007}} && $L_2$ \\ 
 $n$ &&   $\omega_{_{Re}}$   &   $\omega_{_{Im}}$  && $\omega_{_{Re}}$ & $\omega_{_{Im}}$ &&  relative error \\ 
\mr
 0 && 0.49984 & -0.50002 && 0.50000 & -0.50000 && 0.02270\%\\ 
 1 && 0.49961 & -1.49877 && 0.50000 & -1.50000 && 0.08174\%\\ 
 2 && 0.49807 & -2.49714 && 0.50000 & -2.50000 && 0.13540\%\\ 
 3 && 5.01544 & -3.53140 && 0.50000 & -3.50000 && 0.88922\%\\
\br
\end{tabular}
\end{center}
\end{table}

\section{Conclusion}

This proceeding has investigated the accuracy of PINNs in solving perturbation equations with effective potentials expressed in the form of inverted P\"{o}schl-Teller potentials. This work was preliminary to the larger goal of our project, which is to implement PINNs for differential equations of more general black hole perturbation scenarios. So far, our results indicate that PINNs can accurately compute QNFs, for a readily implementable set of hyperparameters. Further work on more general space-times will entail fine-tuning our choice of hyperparameters using a grid-search algorithm similar to the technique employed in Ref.~\cite{Kadeethumetal2020}. Increasing the performance of our PINN models may ultimately require us overcoming the spectral bias that is inherent in fully connected neural networks, an attribute that renders them less accurate for higher frequency solution functions \cite{Wangetal2020}.

\ack
AMN is supported by the Faculty of Science at the University of Johannesburg. GEH was supported by the GES, and ASC is partially supported by the National Research Foundation South Africa.

\end{document}